\documentclass[12pt]{article}
\usepackage{amsfonts}
\def\cA{{\cal A}}
\newcommand{\half}{\frac{1}{2}}
\newcommand{\Half}{\frac{3}{2}}
\begin{document}  
 
\begin{center}
{\Large \textbf{A minimum principle in mRNA editing ?}}

\vspace{10mm}

{\large L. Frappat$^{ac}$, A. Sciarrino$^{b}$, P. Sorba$^{a}$}

\vspace{10mm}

\emph{$^a$ Laboratoire d'Annecy-le-Vieux de Physique Th{\'e}orique LAPTH}

\emph{CNRS, UMR 5108, associ{\'e}e {\`a} l'Universit{\'e} de Savoie}

\emph{BP 110, F-74941 Annecy-le-Vieux Cedex, France}

\vspace{7mm}

\emph{$^b$ Dipartimento di Scienze Fisiche, Universit{\`a} di Napoli 
``Federico II''}

\emph{and I.N.F.N., Sezione di Napoli}

\emph{Complesso Universitario di Monte S. Angelo, Via Cintia, I-80126 
Napoli, Italy}

\vspace{7mm}

\emph{$^c$ Member of the Institut Universitaire de France}

\end{center}

\vspace{12mm}

\begin{abstract}
mRNA editing of sequences of many species is analyzed.  The nature of the 
inserted nucleotides and the position of the insertion sites, once fixed 
the edited peptide chain, are explained by introducing a minimum principle 
in the framework of the crystal basis model of the genetic code introduced 
by the authors.
\end{abstract}

\vfill


\vspace*{3mm}

\rightline{LAPTH-812/00}
\rightline{physics/0009063}


\newpage

\baselineskip=18pt

\section{Introduction}

One of the basic dogmas of molecular genetics states that the information 
contained in DNA flows faithfully, via the mRNA intermediate molecule, into 
the production of proteins.  In 1986 \cite{Ben}, it has been discovered in 
trypanosoma mitochondria that the information contained in DNA is not 
always found unmodified in the RNA products.  In the following fifteen 
years, it has been demonstrated that in several organisms (kinetoplastid 
protozoa, mitochondria or chloroplasts of plants, mammalian cells), some 
yet unknown biochemical machinery alters the sequence of the final 
transcription products.  This process is called \emph{RNA editing}.  For an 
extensive list of articles on RNA editing, the reader can look at the many 
web sites on RNA editing \cite{adresses,Ubiblio}.

The alteration of the sequence of nucleotides in the RNA occurs after it 
has been transcribed from DNA but before it is translated into protein.  
Post-transcriptional modifications have also been observed and interpreted 
as RNA editing.  RNA editing occurs by two distinct mechanisms: 1) 
\emph{substitution editing:} chemical alteration of individual nucleotides 
(the equivalent of point mutations), usually C $\to$ U. These alterations 
are catalyzed by proteins that recognize a specific target sequence of 
nucleotides (much like restriction enzymes).  2) \emph{insertion/deletion 
editing:} insertion or deletion of nucleotides in the RNA (usually U or C).  
It is generally believed that these alterations are mediated by guide RNA 
molecules (gRNA) that base-pair as best they can with the RNA to be edited 
and serve as a template for the addition (or removal) of nucleotides in the 
target \cite{BBS}.  However there is no evidence for the presence of the 
gRNA for all concerned biological species.

The main features of mRNA editing are: \\
-- the insertion (generally multiple) of U nucleotides or of a single C 
nucleotide.  \\
-- the large majority of the transition involves C $\to$ U. A few cases of 
transitions U $\to$ C have also been reported.  \\
-- mRNA editing modifies a few percent (0.8 to 5.8 \%) of the nucleotides 
of a specific transcript.  \\
-- the mRNA editing appears as a random event, but most of the edited 
nucleotides occurs at certain hotspots.

As a consequence of the RNA editing, there is a change in the final 
biosynthesis of amino acids, the most frequent changes being Pro $\to$ Ser, 
Ser $\to$ Leu, Ser $\to$ Phe.  The deep mechanism which causes RNA editing 
is still unknown.  The understanding of the event is complicated: from a 
thermodynamics point of view a change, i.e. C $\to$ U, takes place if it is 
favored in the change of entalpy or entropy, but should this be the case, 
the change should appear in all the organisms.  Moreover from a microscopic 
(quantum mechanical) point of view, the change should occur in both 
directions, i.e. C $\leftrightarrow$ U. It seems that the primary aim of 
mRNA editing is the evolution and conservation of protein structures, 
creating a meaningful coding sequence specific for a particular amino acid 
sequence.

The purpose of this paper is to propose an effective model to describe the 
RNA editing.  Our model does not explain why, where and in which organisms 
editing happens, but it gives a framework to understand some specific 
features of the phenomenon.  The paper is organized as follows.  In section 
2, we analyze the mRNA editing in Physarum polycephalum.  We first consider 
this biological species for two reasons: the high statistics of the 
available data, and the feature of this editing which is mainly 
characterized by single C insertions, allowing a more detailed and accurate 
analysis.  We show that the existence of preferred sites as well as the 
nature of the insertions can be understood by requiring the minimization of 
a suitable function defined on the codon sequence.  This function can be 
defined as we identify each codon by a set of four half-integer labels.  In 
section 3, we then analyze the generally multiple U insertions occuring in 
kinetoplastid protozoa and we show that also in this case the mRNA editing 
is understood by a similar minimization procedure.  In section 4, we 
discuss briefly the substitution editing.  Finally, we give a few 
conclusions and highlights for future developments.

\section{Insertion editing by C}

The mRNA editing in Physarum polycephalum, discovered in 1991 by R. 
Mahendran, M. Spottswood and D. Miller \cite{Miller}, has been extensively 
studied and it presents the peculiar feature to be characterized mainly by 
C insertions.  Main feature of the RNA editing in Physarum polycephalum is 
that in about 80 \% of the cases the insertion occurs in the third position 
of the codon, the insertion sites are non random and in about 68 \% of the 
cases the C is inserted after a purine-pyrimidine dinucleotide.  Moreover 
no rule for the location of the editing sites has been determined, even if 
the presence of hotspots have been remarked.  We have analyzed three 
published sequences of mRNA editing in portion of the ATP-9 Mitochondrial, 
of mRNA of cytochromes c and b of Physarum polycephalum 
\cite{Miller,GVH,WMM}, showing respectively, 54 insertions of a single C, 
62 insertions (59 single C, 1 single U) and 40 insertions (31 single C, 6 
single U).  As a whole we have analyzed 151 single insertions (144 C and 7 
U) in three published sequences of Physarum polycephalum 
\cite{Miller,GVH,WMM}, remarking that the same amino acid chain could have 
been obtained by insertion of C in a site different from the observed one 
or by insertion of a nucleotide different from C or U.

In the whole of the analyzed sequences we have remarked (inserted C 
nucleotides are underlined):
\begin{enumerate}
\item 
the presence of at least 22 alternative insertion sites for C (15 \% of the 
cases, see Table \ref{tableea}), which would produce the same final amino 
acids, so not altering the protein biosynthesis.  For example, at the 
insertion site 9 of Ref.  \cite{Miller}, the (observed) sequence is 
AC\underline{C} TTA (Thr Leu), while the (unobserved) sequence with 
alternative insertion site may be ACT \underline{C}TA.
\item 
in at least 108 (resp.  98 and 63) of the 144 single C insertions (75 \%, 
resp.  68 \% and 44 \% of the cases, see Table \ref{tableea2}), the same 
final amino acid may have been obtained by a single U (resp.  A and G) 
insertion.  Note that in writing Table \ref{tableea2}, when the insertion 
site is ambiguous, i.e. when the inserted C is next to another C, sometimes 
a shift has been performed.
\end{enumerate}
Moreover, we have to consider the two cases GCC UCU $\to$ GCU ACU -- site 
16$'$ -- and CUU AAA $\to$ UUA AAR -- site 21* -- where C insertion is 
replaced by an A or an R (R = A, G) insertion together with a shift of the 
insertion site.  A similar analysis has been performed for the single U 
insertions.

This implies two natural questions: 1) why the insertion sites are the 
observed ones and not the other ones~? 2) why the C insertion is largely 
preferred~?

In physics when a phenomenon occurs in one fixed way between many possible 
choices, one assumes that some minimum principle has to be satisfied.  The 
simplest example is the straight path of light (in absence of strong 
gravitation fields), corresponding to the shortest path between two point 
in euclidean geometry (the so-called geodesics).  Can we think of the 
existence of a sort of \emph{minimum principle} to explain mRNA editing 
and/or other process in DNA~?  There are several technical and conceptual 
difficulties in this way of tackling the problem.  One should give a 
mathematical modelisation of RNA and identify the sequence by a possibly 
discrete set of variables.  Defining a topological metric space depending 
on discrete variables and introducing on it a variation principle is a hard 
mathematical problem.  Moreover we do not have a priori any theoretical 
guidelines, such as the Hamiltonian and/or Lagrangian formalisms, so we 
must have some good empirical grounding to begin with.  Of course we do not 
expect biological processes to be deterministic, as it is the case in 
classical mechanics; so we have to unite minimum principle, if any, with 
random nature of the events, like in quantum mechanics.  In the present 
note, as a first step, we look for a simple function which would take the 
smallest value in the observed configuration of insertion sites and single 
C insertion, with respect to the configuration with insertion in 
alternative sites and/or with a single U, G, A insertion.

The starting point for a mathematical modelisation of DNA or mRNA is the 
crystal basis model of the genetic code \cite{FSS1} where the nucleotides 
are assigned to the 4-dim irreducible fundamental representation $(1/2, 
1/2)$ of $U_{q \to 0}(sl(2) \oplus sl(2))$ and any sequence of $N$ 
nucleotides to the $N$-fold tensor product of $(1/2, 1/2)$ (for codons, see 
\cite{FSS1} or Table 4 of \cite{FSS2}, here reported in Table 
\ref{tablerep} for completeness).  As a consequence of the model any 
nucleotide sequence is characterized as an element of a vector space.  
Therefore, functions can be defined on this space and can be computed on 
the sequence of codons.  Maybe it is worthwhile to emphasize that for the 
aim of this paper, it is not necessary to undestand completely either the 
mathematical structure of the crystal basis, or the reason to deal with 
such a sophisticated mathematical structure (see e.g. \cite{FSS1,FSS2}).  
The essential point is that any codon is identified by a set of four 
half-integer labels and functions can be defined on the codons.  We make 
the assumption that the location sites for the insertion of a nucleotide 
should minimize the following function for the mRNA or cDNA
\begin{equation}
\cA_{0} = \exp \left[ -\sum_{k} 4 \alpha_{c} \, C_{H}^{k} \, + \, 4 \beta_{c} 
\, C_{V}^{k} \, + \, 2 \gamma_{c} J_{3,H}^{k} \right]
\label{eq:A} 
\end{equation} 
where the sum in $k$ is over all the codons in the edited sequence, 
$C_{H}^{k} $ ($C_{V}^{k} $) and $J_{3,H}^{k}$, are the values of the 
\emph{Casimir} operator and of the third component of the generator of the 
$H$-$sl(2)$ ($V$-$sl(2)$), see \cite{FSS1}, in the irreducible 
representation to which the $k$-th codon belongs, see Table 1.  Let us 
recall that the value of the Casimir operator on a state in an irreducible 
representation (IR) labelled by $(J_{H}, J_{V})$ is
\begin{equation}
C_{H} \, (J_{H}, J_{V}) = J_{H}(J_{H} + 1) \quad \mbox{and} \quad C_{V} \, 
(J_{H}, J_{V}) = J_{V}(J_{V} + 1)
\end{equation}
In (\ref{eq:A}) the simplified assumption that the dependence of ${\cal 
A}_{0}$ on the irreducible representation to which the codon belongs is 
given only by the values of the Casimir operators has been made.  The 
parameters $\alpha_{c}, \beta_{c}, \gamma_{c}$ are constants, depending on 
the biological species.

The minimum of $\cA_{0}$ has to be computed in the whole set of 
configurations satisfying to the constraints: i) the starting point should 
be the mtDNA and ii) the final peptide chain should not be modified.  It is 
obvious that the global minimization of expression (\ref{eq:A}) is ensured 
if $\cA_{0}$ takes the smallest value locally, i.e. in the neighborhood of 
each insertion site.  The form of the function ${\cal A}_{0}$ is rather 
arbitrary; one of the reasons of this choice is that the chosen expression 
is computationally quite easily tractable.  If the parameters $\alpha_{c}, 
\beta_{c}, \gamma_{c}$ are strictly positive with $\gamma_{c}/6 > \beta_{c} 
> \alpha_{c}$, the minimization of (\ref{eq:A}) explains the observed 
configurations in all cases, except for the cases 12, 33, 45 and 41* where 
there is equality and the cases 18* and 51* where the minimization is not 
satisfied (see Table \ref{tableea}).

In order to deal with the remaining cases and to take into account the 
observed fact that the dinucleotide preceding the insertion site is 
predominantly a purine-pyrimidine, we add to the exponent of the function 
$\cA_{0}$ an "interaction term" which is equivalent to multiply 
(\ref{eq:A}) by the function $\cA_{1}$ where
\begin{equation}
\cA_{1} = \exp \left[ \sum_{i} - 4 \, \omega_{1c} \, j_{3,V}^{(i)} 
\cdot j_{3,V}^{(i-1)} + 4 \, \omega_{2c} \, j_{3,V}^{(i)} \cdot 
j_{3,V}^{(i-2)} \right]
\label{eq:A1} 
\end{equation} 
The sum in $i$ is over the insertion sites and $j_{3,V}^{(i-n)}$ is the 
value of the third component of the generator of $V$-$sl(2)$ of the $n$-th 
nucleotide preceding the inserted nucleotide C (i.e. $+1/2$ for C, U and 
$-1/2$ for G, A) and $\omega_{1c}, \omega_{2c}$ are constants, depending on 
the biological species.  In the case where the insertion site cannot be 
unambiguously determined, i.e. when the inserted nucleotide is next to a 
nucleotide of the same type, (\ref{eq:A1}) should be computed in the 
configuration which minimizes the value of $\cA_{1}$.  If $\omega_{1c} > 
\omega_{2c} > 0$ and $\omega_{1c} > 12 \alpha_{c}$ the minimization of the 
function $\cA = \cA_{0} \, \cA_{1}$ explains all the observed positions for 
C insertions, see Table \ref{tableea}.  It is reasonable, but not taken 
into account in (\ref{eq:A}), to argue that the insertion sites and the 
nature of the inserted nucleotides also depend on the content of the 
particular sequence.  Moreover $\cA$ might be considered as the first terms 
of a development, next terms involving representations corresponding to 
more than one codon, the nature of the nucleotides following the insertion 
site, etc.  These further terms may play a role in a more refined analysis.

An analysis of the 7 single U insertions shows that in 6 cases -- sites 
22*, 10$'$, 18$'$, 22$'$, 24$'$, 26$'$ -- (resp.  3 cases -- sites 10$'$, 
18$'$, $24'$ --) the replacement U $\to$ C (resp.  U $\to$ R) gives the 
same amino acid.  In 4 of these cases the minimization of Eq.  (\ref{eq:A}) 
should prefer the insertion of C, giving rise to UUU $\to$ UUC, site 22*; 
CUU $\to$ CUC, sites 10$'$, 18$'$; ACU $\to$ ACC, site 24$'$, while in 
sites 22$'$, 26$'$ UUA is more preferred than CUA. This may explain why the 
U insertions are so rare compared with the C insertions.  Also in this case 
further terms in $\cA$ may help for a more refined analysis of the 
preferred configuration of the insertions.

\section{Insertion editing by U}

The mRNA editing with insertion of U has been observed in particular in a 
group of parasitic protozoa known as kinetoplastid protozoa.  Contrary to 
the C insertion case where only single nucleotide insertions occurs, the 
main characteristics of the mRNA editing by U insertion is that the U 
nucleotides are inserted by blocks.  In this way, almost all amino acids 
are can be obtained with a great proportion of Phe and Leu.  Many sequences 
where mRNA editing with U insertion occur can be found in \cite{Udata} and 
an extensive list of references on the U insertion editing can be found in 
\cite{Ubiblio}.  We limit ourselves to cite the first papers on the subject 
\cite{SFSS1988,FSSS1988,FAS,SS1989}.  The table below shows the species and 
the genes that have been used in our analysis.  In this table, COX = 
cytochrome oxidase, Cyt b = cytochrome b, G = G-rich region, NADH = NADH 
dehydrogenase, RPS12 = ribosomal protein S12, MURF = maxicircle 
unidentified reading frame.  The number of edited sites is quite large 
(more than 1000 sites).

{\small \begin{center}
\begin{tabular}{|l|ccccccc|}
\hline
species & ATPase6 & COX I & COX II & COX III & Cyt b & G3 & G4 \\
\hline
Crithiadia fasciculata & X & & X & X & X & & \\
Leishmania tarentolae & X & & X & X & X & X & X \\
Phytomonas serpens & X & & & & & X & \\
Trypanosoma brucei & X & & & X & X & & X \\
Trypanosoma borreli & & X & & & X & & \\
Trypanosoma cruzi & X & & X & & & & \\
\hline
species & MURF2 & NADH3 & NADH7 & NADH8 & NADH9 & RPS12 & \\
\hline
Crithiadia fasciculata & X & & X & & & X & \\
Leishmania tarentolae & X & X & X & X & X & X & \\
Phytomonas serpens & & & & X & & X & \\
Trypanosoma brucei & X & X & X & X & X & X & \\
Trypanosoma borreli & & & & & & X & \\
Trypanosoma cruzi & & & & & & & \\
\hline
\end{tabular}

\vspace*{3mm}

Species and genes used in the U insertion mRNA editing analysis.
\end{center}
}

Following the same analysis as in the previous section, we make the 
assumption that the location sites for the insertion of a U nuleotide 
should minimize the following function for the mRNA:
\begin{equation}
\cA'_{0} = \exp \left[ -\sum_{k} 4 \alpha_{u} \, C_{H}^{k} \, + \, 4 
\beta_{u} \, C_{V}^{k} \, + \, 2 \gamma_{u} J_{3,H}^{k} \right]
\label{eq:A0} 
\end{equation} 
When choosing the parameters $\alpha_{u}$, $\beta_{u}$, $\gamma_{u}$ such 
that $\alpha_{u},\gamma_{u} < 0$ and $\beta_{u} > 0$ with $\gamma_{u}/6 < 
\alpha_{u}$, the minimization of (\ref{eq:A0}) explains all the observed 
configuration, except in the cases CG\underline{U} and GG\underline{U} 
where the configurations CGA and CG\underline{U} on the one hand and GGA 
and GG\underline{U} on the other hand are equivalent (inserted U are 
underlined).
Multiplying Eq. (\ref{eq:A0}) by the corrective term
\begin{equation}
\cA'_{1} = \exp \left[ \sum_{i} - 4 \, \omega_{1u} \, j_{3,V}^{(i)} 
\cdot j_{3,V}^{(i-1)} \right]
\label{eq:A1U} 
\end{equation} 
with $\omega_{1u} < 0$, the observed configurations become the preferred 
ones.

It may happen that different U insertions lead to the same configuration of 
amino-acids (note however that in the U insertion case, this is much less 
frequent than in the C insertion case, since in the U insertion case, the U 
nucleotides are inserted by blocks).  In the analyzed sequences, we have 
noticed six such possible alternative configurations: \\
-- in Leishmania tarentola, gene NADH8, at edited position 229, one 
observes the configuration $C_{1} = $ GC\underline{U} CUA, the alternative 
configuration is $C_{2} = $ GCC \underline{U}UA, and one has 
$\cA_{0}(C_{1}) < \cA_{0}(C_{2})$.  \\
-- in Phytomonas serpens, gene NADH8, at edited position 355, one observes 
the configuration $C_{1} = $ GCA A\underline{UU}, the alternative 
configuration is $C_{2} = $ GC\underline{U} A\underline{U}A, and both 
configurations are equivalent: ${\cal A}_{0}(C_{1}) = \cA_{0}(C_{2})$.  \\
-- in Trypanosoma borreli, gene cytochrome c oxidase I, at edited position 
1375, one observes the configuration $C_{1} = $ G\underline{U}A 
A\underline{UU}, the alternative configuration is $C_{2} = $ 
G\underline{UU} A\underline{U}A, both configurations are equivalent: 
$\cA_{0}(C_{1}) = \cA_{0}(C_{2})$.  \\
-- in Trypanosoma brucei, gene cytochrome oxidase III, at edited position 
645, one observes the configuration $C_{1} = $ GCA \underline{UU}G 
\underline{UU}A \underline{UUU} A\underline{UU}, the alternative 
configuration is $C_{2} = $ GC\underline{U} \underline{UU}A \underline{UU}G 
\underline{UUU} A\underline{U}A, both configurations are equivalent: 
$\cA_{0}(C_{1}) = \cA_{0}(C_{2})$.  \\
-- in Trypanosoma brucei, gene NADH7, at edited position 988, one observes 
the configuration $C_{1} = $ CCG GG\underline{U}, the alternative 
configuration is $C_{2} = $ CC\underline{U} GGG, and one has 
$\cA_{0}(C_{1}) > {\cal A}_{0}(C_{2})$.  This is a counter-example, however 
in the configuration $C_{2}$, the nucleotide U is inserted after a C, which 
is not favored.  \\
-- in Trypanosoma brucei, gene NADH8, at edited position 251, one observes 
the configuration $C_{1} = $ \underline{U}GC CC\underline{U}, the 
alternative configuration is $C_{2} = $ \underline{U}G\underline{U} CCC, 
and one has $\cA_{0}(C_{1}) > {\cal A}_{0}(C_{2})$.  This one is also a 
counter-example.

In the above cases where the insertion sites are not unambiguously 
determined, multiplying Eq. (\ref{eq:A0}) by the following corrective term
\begin{equation}
\cA''_{1} = \exp \left[ \sum_{i} - 4 \, \omega_{1u} \, j_{3,V}^{(i)} 
\cdot j_{3,V}^{(i-1)} + 4 \, \omega_{2u} \, j_{3,V}^{(i)} \cdot 
j_{3,V}^{(i-2)} \right]
\label{eq:A1sec} 
\end{equation} 
with $\omega_{1u} < 0$ and $\omega_{1u} + \omega_{2u} > 0$, the observed 
configurations become the preferred ones.

In conclusion, the observed U insertions minimize the function $\cA' = 
\cA'_{0} \cA''_{1}$, except for two cases for which alternative insertion 
sites exist, where the function $\cA'$ takes a lower value, at least in the 
simplified hypothesis that $\cA''_{1}$ is a perturbative term to 
$\cA'_{0}$.  It should however be noted that such perturbative term takes 
into account the nature of the neighbor nucleotides and the experimentally 
observed bias in the selection of the insertion sites shows an important 
effect of the neighbors.

\section{Substitution editing by C $\to$ U}

Substitution editing of mRNA by C $\to$ U occurs for example in plant 
mitochondria and chloroplasts and in the gene apoB in mammals (see web site 
\# 4 in Ref.  \cite{adresses}).  For our study we have used the COXII gene 
of the wheat \cite{Covello}.  Similar radical amino acid substitutions in 
plant COXII sequences have been inferred.  Although the statistics is 
rather poor, we can extract interesting features.  In the wheat COXII gene, 
one observes the following substitutions: CGG $\to$ UGG (twice), CCU $\to$ 
UCU, UCA $\to$ UUA (twice), UCG $\to$ UUG, CGU $\to$ UGU, ACG $\to$ AUG, so 
that the corresponding amino acids Trp, Leu, Leu, Cys, Met are correctly 
coded by the universal code.  In \cite{wheat}, the following substitution 
editing has been observed in several wheat genes (COXII, COXIII, Cob, NAD3, 
NAD4, RPS12): CGG $\to$ UGG (seven times), CAC $\to$ UAC, CAU $\to$ UAU, 
UCA $\to$ UUA, UCG $\to$ UUG (twice), UCU $\to$ UUU (three times), CUC 
$\to$ UUC, CCG $\to$ CUG, CCA $\to$ CUA. In the case of the mammalian gene 
apoB, the editing depends on the location of the mRNA in the body of the 
species under consideration (editing in the intestine but no editing in the 
liver). It is characterized by CAA $\to$ UAA (Gln $\to$ Stop codon).

As before, one can easily check that the function $\cA'_{0}$ of Eq.  
(\ref{eq:A0}) minimizes the configuration corresponding to the substituted 
nucleotide with respect to the original one.

In \cite{GVH}, three cases of substitution editing are reported in the coI 
gene of Physarum polycephalum.  Also in this the function $\cA'_{0}$ is 
minimized.  However, this function differs from the function $\cA_{0}$ Eq.  
\ref{eq:A} of Physarum.

\section{Conclusion}

We have shown that the nature of the inserted nucleotides and the position 
of the insertion site can be explained by introducing a minimum principle 
in the framework of the crystal basis model of the genetic code introduced 
in ref.  \cite{FSS1}.  Indeed, we have made the assumption that, once fixed 
the final edited peptide chain, the nature and the position of the inserted 
nucleotide(s), are such to minimize the functions eqs.  
(\ref{eq:A})-(\ref{eq:A1}) or (\ref{eq:A0}), where the numerical real 
coefficients depend on the biological species, and the operators $C_{H,V}$ 
and $J_{3,H}$ have to be evaluated on the edited codons using Table 
\ref{tablerep}.

Our analysis shows that, in the case of Physarum polycephalum, in 110 of 
the 114 sites in which the insertion of C or U, and in all the cases where 
also an insertion of purine can produce the same amino acid, the observed 
mRNA editing makes use of the nucleotide C or U which does minimize $\cA = 
\cA_{0} \cA_{1}$.  In the case of the U insertion in kinetoplastid protozoa 
genes, in all the cases but two, the function $\cA'$ is minimized.  This 
last function is also minimized in the case of C $\to$ U substitution 
editing.

The form of the function assumed to be minimized has been suggested by 
simplicity and easiness of computation.  For these reasons we have only 
considered a dependence on the values of the Casimir operator $C_{H}$ and 
$C_{V}$, although generally there is a degeneracy in the irreducible 
representations.  We have also made the hypothesis that the effects of 
neighboring nucleotides is weak and limited to the two foregoing ones.  As 
we said previously, we are first of all investigating solid empirical 
grounds bearing the approach under consideration out, looking then for 
further mathematical refinements which may give also quantitative 
information.

We have not considered insertion by nucleotides different from C and U 
since the statistics is very low.  We have assumed that the constants 
$\alpha,\beta,\gamma$ depend on the biological species.  However our 
analysis cannot exclude that indeed they depend only on the type of the 
inserted nucleotide.  It would be interesting to analyze further data on 
mRNA editing in the analyzed as well as in other biological species to 
check that the minimum principle is satisfied.  Further confirmation of the 
validity of our hypothesis would provide evidence in favor of the existence 
of strong physical chemical constraints in the domain generally believed 
dominated by casual events.  The presence of a minimum principle which is 
indeed an indication of the possible application of variational principle 
in the field of complex biological systems would be an amazing result.

In conclusion our \emph{effective model} does not explain why and where 
mRNA editing occurs, but it seems to be able to determine the location 
sites and the nature of inserted nucleotides, once fixed the amino acid 
chain.

\bigskip

\textbf{Acknowledgements}: It is a pleasure to thank D.L.~Miller and 
L.~Simpson for pointing us the references on RNA editing.  Partially 
supported by MURST (Italy) and MAE (France) in the framework of 
french-italian collaboration Galileo.

\clearpage

\begin{table}[htbp]
\caption{The eukariotic code. The upper label denotes different irreducible
representations.}
\label{tablerep}
\footnotesize
\begin{center}
\begin{tabular}{|cc|cc|rr|cc|cc|rr|}
\hline
codon & a.a. & $J_{H}$ & $J_{V}$ & $J_{3,H}$ & $J_{3,V}$& codon & a.a. & 
$J_{H}$ & $J_{V}$ & $J_{3,H}$ & $J_{3,V}$ \\
\hline
\Big. CCC & Pro & $\Half$ & $\Half$ & $\Half$ & $\Half$ & UCC & Ser & $\Half$ & 
$\Half$ & $\half$ & $\Half$ \\
\Big. CCU & Pro & $(\half$ & $\Half)^1$ & $\half$ & $\Half$ & UCU & Ser & 
$(\half$ & $\Half)^1$ & $-\half$ & $\Half$ \\
\Big. CCG & Pro & $(\Half$ & $\half)^1$ & $\Half$ & $\half$ & UCG & Ser & 
$(\Half$ & $\half)^1$ & $\half$ & $\half$ \\
\Big. CCA & Pro & $(\half$ & $\half)^1$ & $\half$ & $\half$ & UCA & Ser & 
$(\half$ & $\half)^1$ & $-\half$ & $\half$ \\[1mm]
\hline
\Big. CUC & Leu & $(\half$ & $\Half)^2$ & $\half$ & $\Half$ & UUC & Phe & $\Half$ 
& $\Half$ & $-\half$ & $\Half$ \\
\Big. CUU & Leu & $(\half$ & $\Half)^2$ & $-\half$ & $\Half$ & UUU & Phe & $\Half$ 
& $\Half$ & $-\Half$ & $\Half$ \\
\Big. CUG & Leu & $(\half$ & $\half)^3$ & $\half$ & $\half$ & UUG & Leu & 
$(\Half$ & $\half)^1$ & $-\half$ & $\half$ \\
\Big. CUA & Leu & $(\half$ & $\half)^3$ & $-\half$ & $\half$ & UUA & Leu & 
$(\Half$ & $\half)^1$ & $-\Half$ & $\half$ \\[1mm]
\hline
\Big. CGC & Arg & $(\Half$ & $\half)^2$ & $\Half$ & $\half$ & UGC & Cys & 
$(\Half$ & $\half)^2$ & $\half$ & $\half$ \\
\Big. CGU & Arg & $(\half$ & $\half)^2$ & $\half$ & $\half$ & UGU & Cys & 
$(\half$ & $\half)^2$ & $-\half$ & $\half$ \\
\Big. CGG & Arg & $(\Half$ & $\half)^2$ & $\Half$ & $-\half$ & UGG & Trp & 
$(\Half$ & $\half)^2$ & $\half$ & $-\half$ \\
\Big. CGA & Arg & $(\half$ & $\half)^2$ & $\half$ & $-\half$ & UGA & Ter & 
$(\half$ & $\half)^2$ & $-\half$ & $-\half$ \\[1mm]
\hline
\Big. CAC & His & $(\half$ & $\half)^4$ & $\half$ & $\half$ & UAC & Tyr & 
$(\Half$ & $\half)^2$ & $-\half$ & $\half$ \\
\Big. CAU & His & $(\half$ & $\half)^4$ & $-\half$ & $\half$ & UAU & Tyr & 
$(\Half$ & $\half)^2$ & $-\Half$ & $\half$ \\
\Big. CAG & Gln & $(\half$ & $\half)^4$ & $\half$ & $-\half$ & UAG & Ter & 
$(\Half$ & $\half)^2$ & $-\half$ & $-\half$ \\
\Big. CAA & Gln & $(\half$ & $\half)^4$ & $-\half$ & $-\half$ & UAA & Ter & 
$(\Half$ & $\half)^2$ & $-\Half$ & $-\half$ \\[1mm]
\hline
\Big. GCC & Ala & $\Half$ & $\Half$ & $\Half$ & $\half$ & ACC & Thr & $\Half$ & 
$\Half$ & $\half$ & $\half$ \\
\Big. GCU & Ala & $(\half$ & $\Half)^1$ & $\half$ & $\half$ & ACU & Thr & 
$(\half$ & $\Half)^1$ & $-\half$ & $\half$ \\
\Big. GCG & Ala & $(\Half$ & $\half)^1$ & $\Half$ & $-\half$ & ACG & Thr & 
$(\Half$ & $\half)^1$ & $\half$ & $-\half$ \\
\Big. GCA & Ala & $(\half$ & $\half)^1$ & $\half$ & $-\half$ & ACA & Thr & 
$(\half$ & $\half)^1$ & $-\half$ & $-\half$ \\[1mm]
\hline
\Big. GUC & Val & $(\half$ & $\Half)^2$ & $\half$ & $\half$ & AUC & Ile & $\Half$ 
& $\Half$ & $-\half$ & $\half$ \\
\Big. GUU & Val & $(\half$ & $\Half)^2$ & $-\half$ & $\half$ & AUU & Ile & $\Half$ 
& $\Half$ & $-\Half$ & $\half$ \\
\Big. GUG & Val & $(\half$ & $\half)^3$ & $\half$ & $-\half$ & AUG & Met & 
$(\Half$ & $\half)^1$ & $-\half$ & $-\half$ \\
\Big. GUA & Val & $(\half$ & $\half)^3$ & $-\half$ & $-\half$ & AUA & Ile & 
$(\Half$ & $\half)^1$ & $-\Half$ & $-\half$ \\[1mm]
\hline
\Big. GGC & Gly & $\Half$ & $\Half$ & $\Half$ & $-\half$ & AGC & Ser & $\Half$ & 
$\Half$ & $\half$ & $-\half$ \\
\Big. GGU & Gly & $(\half$ & $\Half)^1$ & $\half$ & $-\half$ & AGU & Ser & 
$(\half$ & $\Half)^1$ & $-\half$ & $-\half$ \\
\Big. GGG & Gly & $\Half$ & $\Half$ & $\Half$ & $-\Half$ & AGG & Arg & $\Half$ & 
$\Half$ & $\half$ & $-\Half$ \\
\Big. GGA & Gly & $(\half$ & $\Half)^1$ & $\half$ & $-\Half$ & AGA & Arg & 
$(\half$ & $\Half)^1$ & $-\half$ & $-\Half$ \\[1mm]
\hline
\Big. GAC & Asp & $(\half$ & $\Half)^2$ & $\half$ & $-\half$ & AAC & Asn & $\Half$ 
& $\Half$ & $-\half$ & $-\half$ \\
\Big. GAU & Asp & $(\half$ & $\Half)^2$ & $-\half$ & $-\half$ & AAU & Asn & $\Half$ 
& $\Half$ & $-\Half$ & $-\half$ \\
\Big. GAG & Glu & $(\half$ & $\Half)^2$ & $\half$ & $-\Half$ & AAG & Lys & $\Half$ 
& $\Half$ & $-\half$ & $-\Half$ \\
\Big. GAA & Glu & $(\half$ & $\Half)^2$ & $-\half$ & $-\Half$ & AAA & Lys & $\Half$ 
& $\Half$ & $-\Half$ & $-\Half$ \\[1mm]
\hline
\end{tabular}
\end{center}
\end{table}

\newpage 

\begin{table}[t]
\caption{ From the left: the a.a., the C insertion site, the codons coding 
for the a.a., the dinucleotide preceding C; the shift with respect to the 
observed site of the alternative insertion site, the new codons, the 
dinucleotide preceding C in the alternative site.  Ref.  to fig.  3 of 
\cite{Miller}, fig.  2 of \cite{GVH} (with an asterisk *), fig.  2 of 
\cite{WMM} (with a prime $'$).}

\label{tableea}
\footnotesize
\begin{center}
\begin{tabular}{|c||c|c|c||c|c|c|}
\hline
a.a. & site & codons & dinucl. & shift  & codons & dinucl. \\
\hline
& & & & & & \\
Thr, Leu & 9, 24, 55* & ACC, UUA & AC & + 1 & ACU, CUA & CU \\
& & & & & & \\
\hline
& & & & & & \\
Ile, Leu & 23, 30* & AUC, UUG & AU & $+1$ & AUU, CUG & UU \\
& & & & & & \\
\hline
& & & & & & \\
Ala, Phe & 32 & GCC, UUU & GC & $+3$ & GCU, UUC & UU \\
& & & & & & \\
\hline
& & & & & & \\
Val, Phe & 33, 45, 41* & GUC, UUU & GU & $+3$ & GUU, UUC & UU \\
& & & & & & \\
\hline
& & & & & & \\
Ser, Arg & 34 & UCC, AGA & UC & $+1$ & UCA, CGA & CA \\
& & & & & & \\
\hline
& & & & & & \\
Asn, Phe & 12 & AAU, UUC & UU & $-3$ & AAC, UUU & AA \\
& & & & & & \\
\hline
& & & & & & \\
Ile, Leu & 49, 48*, 20$'$ & AUC, UUA & AU & $+1$ & AUU, CUA & UU \\
& & & & & & \\
\hline
& & & & & & \\
Ala, Leu & 5$'$ & GCC, UUA & GC & $+1$ & GCU, CUA & CU \\
& & & & & & \\
\hline
& & & & & & \\
Ser, Phe & 43*, 13$'$ & UCC, UUU & UC & $+3$ & UCU, UUC & UU \\
& & & & & & \\
\hline
& & & & & & \\
Thr, Arg & 3* & ACC, AGA & AC & $+1$ & ACA, CGA & CA \\
& & & & & & \\
\hline
& & & & & & \\
Ser, Leu & 18* & AGU, CUG & GU & $-1$ & AGC, UUG & AG \\
& & & & & & \\
\hline
& & & & & & \\
Val, Leu & 23*, 40* & GUC, UUA & GU & $+1$ & GUU, CUA & UU \\
& & & & & & \\
\hline
& & & & & & \\
His, Leu & 51* & CAU, CUA & AU & $-1$ & CAC, UUA & CA \\
& & & & & & \\
\hline
\hline
\end{tabular}
\end{center}
\end{table}

\newpage

\begin{table}[t]
\caption{ From the left: the a.a., the codon created by C insertion, the 
alternative codon created by alternative insertion, the site with reference 
to fig.  3 of \cite{Miller}, fig.  2 of \cite{GVH} (with an asterisk *), 
fig.  2 of \cite{WMM} (with a prime $'$).  Here X = U, A, G and R = A, G.}
\label{tableea2}
\footnotesize
\begin{center}
\begin{tabular}{|c||c||c||c|}
\hline
a.a. & codon & alt.  codon & site \\
\hline
& & & \\
Asn & AAC & AAU & 35, 4$'$ \\
& & & \\
Thr & ACC & ACX & 5, 7, 9, 10, 21, 24, 26, 36, 3*, 4*, 5*, 12*, 20*, 26* \\
& & & 33*, 35*, 39*, 49*, 50*, 55*, 62*, 15$'$, 39$'$ \\
& & & \\
Ser & AGC & AGU & 1*, 36*, 34$'$ \\
& & & \\
Ile & AUC & AUU, AUA & 1, 4, 13, 15, 17, 18, 20, 23, 38, 46, 49, 50, 51, 6* \\
& & & 7*, 9*, 16*, 17*, 19*, 24*, 27*, 30*, 34*, 38* \\
& & & 48*, 54*, 57*, 58*, 60*, 61*, 20$'$, 32$'$, 36$'$, 37$'$ \\
& & & \\
His & CAC & CAU & 44* \\
& & & \\
Pro & CCC & CCX & 17$'$ \\
& & & \\
Arg & CGA & AGA & 30 \\
& & & \\
Leu & CUA & UUA & 31, 40, 8*, 51*, 6$'$ \\
& & & \\
Leu & CUG & UUG & 18* \\
& & & \\
Leu & CUC & CUX & 22 \\
& & & \\
Leu & CUU & UUR & 3, 13*, 21*, 47*, 8$'$, 39$'$ \\
& & & \\
Asp & GAC & GAU & 54 \\
& & & \\
Ala & GCC & GCX & 25, 27, 29, 32, 37, 10*, 13*, 28*, 53*, 5$'$, 16$'$, 
27$'$, 30$'$ \\
& & & \\
Val & GUC & GUX & 2, 6, 11, 14, 33, 42, 45, 23*, 40*, 41*, 56*, 9$'$, 
21$'$, 25$'$ \\
& & & \\
Tyr & UAC & UAU & 43 \\
& & & \\
Ser & UCC & UCX & 34, 42*, 2$'$, 12$'$, 13$'$ \\
& & & \\
Phe & UUC & UUU & 12, 52, 45* \\
& & & \\
\hline
\hline
\end{tabular}
\end{center}
\end{table}

\end{document}